\documentclass[pdflatex]{article}
\usepackage{amssymb,amsmath,amsfonts}
\usepackage{scalerel}
\usepackage{graphics,graphicx} % Required for inserting images
\usepackage{tikz} % automata
\usetikzlibrary{automata,positioning}
\usetikzlibrary{patterns,decorations.pathreplacing,decorations.markings}
\usepackage[latin1]{inputenc}
\usepackage{yfonts}
\usepackage{proof}
\usepackage{array}
\def \A {{\mathcal{A}}}

\def \C {{\mathcal{C}}}
\def \D {{\mathcal{D}}}
\def \E {{\mathcal{E}}}

\def \K {{\mathcal{K}}}

\def \M {{\mathcal{M}}}

\def \o {{\mathcal{O}}}

\def \S {{\mathcal{S}}}
\def \T {{\mathcal{T}}}

\def \W{{\mathcal{W}}}

\def \t {{\mathfrak t}}
\newtheorem{defn}{Definition}

\newtheorem{lemma}{Lemma}
\newtheorem{propn}{Proposition}

\def \lft {\noindent}
\def \ovr {\overline}
\def \otau {\overline{\tau}}

\newcommand{\ignore}[1]{}
\newcommand\Tau{\scalerel*{\tau}{T}\!\!}

\newcommand{\brho}{\mathbf{f}}
\newcommand{\disp}[1]{\vspace*{-1em}
    \begin{center} {#1} \end{center} \vspace*{-1mm} }

\date{}

\begin{document}
 
\title{{\Large Distributed  Transition System with Tags and  \\
           Value-wise Metric, for Privacy Analysis} }

%\maketitle

\author{ \small 
             Siva Anantharaman$^1$  \and
             Sabine Frittella$^2\footnote{The research of Sabine Frittella was funded by the grant ANR JCJC 2019, project PRELAP (ANR-19-CE48-0006)}$  \and
             Benjamin Nguyen$^3$ 
   \  \\   \small
   $^1$ LIFO, Universit\'e d'Orl\'eans (France), 
                                email: {\tt siva@univ-orleans.fr} \\
   \small 
   $^2$ INSA-CVL,  LIFO, Universit\'e d'Orl\'eans (France), 
                                  email: {\tt sabine.frittella@insa-cvl.fr} \\
   \small
    $^3$ INSA-CVL,  LIFO, Universit\'e d'Orl\'eans (France), 
                       email: {\tt benjamin.nguyen@insa-cvl.fr}  }

\maketitle

 \begin{abstract}
   \scriptsize{
  We introduce a logical framework  named {\em Distributed  Labeled Tagged Transition 
    System} (DLTTS), using concepts from Probabilistic Automata, Probabilistic  Concurrent
  Systems,  and Probabilistic  labelled transition systems.   We show that  DLTTS can be
  used to formally model how a given piece  of {\em private} information $P$ (e.g., a
  set of tuples) stored in a  given  database   $D$  can get captured progressively by
  an  adversary $A$ repeatedly   querying  $D$,   enhancing the knowledge acquired  
  from the answers to these queries   with   relational deductions using certain additional
  non-private data.  The database $D$ is assumed protected with generalization  mechanisms.
  We also show that, on a large class of  databases, metrics can be  defined `value-wise',  
  and more general notions of adjacency  between data bases can be defined, based on
  these metrics.  These notions  can also play a role in differentially private  protection
  mechanisms.  }

    \vspace*{2mm}\lft

 \end{abstract}
 \section{ { \large Introduction} }
 
 Data anonymization has been investigated for decades, and many privacy
models have been proposed ($k$-anonymity, \emph{differential privacy}, \ldots)
whose goals are to protect sensitive information. In this paper, our goal is not to
define a new privacy  model, but rather to propose a logical framework (agnostic
to the privacy model) to formally
model how the information stored in a database can get captured progressively by
any agent repeatedly querying the database. This model can also be used to quantify
reidentification attacks on a database. We start  with the observation that
databases  are  distributed over several `worlds' in general, and querying such
databases leads to answers which would also be distributed; and conceivably, to such
distributed  answers  one could assign probability distributions  of pertinence to  the
query.   We present a  logical framework that formally  models how a given
{\em private}  information $P$, stored on a  given (distributed) database $D$,
can get captured  progressively   by an  adversary  querying  the   database
repeatedly.  Named DLTTS   ({\em Distributed  Labeled Tagged Transition System}),
the  framework borrows  ideas  from  several domains, such as:
  Probabilistic Automata  of  Segala, Probabilistic  Concurrent  Systems,  and
  Probabilistic  labelled transition systems.
  To every node on a DLTTS will be attached a  {\em tag} (which we will formally
  define later) representing the `current'
  knowledge of the adversary, acquired from the responses  to his/her queries by
  the answering mechanism, at the nodes  traversed  on a run;   this  knowledge can
  be `saturated'  by the adversary  at any node on the run,   using (a given class of)
  relational  deductions  in combination  with public  information   from some  external
  databases, given in  advance.
  
  In our developments below, the  data in the databases will all be  with finite
  domain,  of any  of the following `basic'  types: numerical, non-numerical,
  or  literal. Some data could be structured in a complex taxonomical relation
  (e.g. an ontology), but  we shall only consider simple tree-structured taxonomies
  in this work.  Part of  the data  could also be  `anonymized' via a generalization
  mechanism (e.g. finite intervals or   finite  sets, instead of precise values), over
  the   basic types.  We therefore consider the   types of  the data  in such an
  extended/overloaded sense. (cf. Example 1 below) 
  The paper is structured as follows:
  
   \vspace*{1mm}
    {%\small
     
   % \lft
   $\bullet$ Preliminaries and notations are presented  in Section~\ref{prelim}. 

  %\lft
  $\bullet$ DLTTS are defined formally in Section~\ref{DLTTS}. They are meant as 
  models for how the information stored in a (distributed) database $D$ can get captured
  progressively, by an adversary querying repeatedly  that database.
  We introduce a `blackbox'  mechanism as an `auxiliary' to a DLTTS; denoted as $\o$,
  it will  be conveniently  referred to  as an  oracle. A motivating and  illustrative `running'
  example is given in subsection~\ref{RunningExample}, 

  %\lft
  $\bullet$ A `value-based' distance function $\rho$ is defined in   Section~\ref{DataWise}, 
  first between data of `compatible types' in a natural manner, and subsequently extended
  as a (partial) distance function between `compatible sets' of data tuples.   $\rho$ is
  assumed known only to the oracle $\o$, to which can be assigned a role of  control  on
  the runs of adversary $A$:  At any node along a run of $A$ on the DLTTS, where the 
  saturated knowledge of $A$ gets too $\rho$-close to private/secret  data in $D$ (in
  a sense made  precise in Section~\ref{DataWise}),   force  the run to terminate.
  (At any other node, $\o$ would indicate possible transitions with their  distributions;
  for $A$ to continue the run at his/her own choice.)

 %\lft
  $\bullet$ The notions of $\epsilon$-distinguishability,  $\epsilon$-LDP,  and 
  $\epsilon$-DP for databases are recalled in Sections~\ref{LDP} and \ref{DP};   and they
  are refined in combination with a vision  based on metrics,  in Section~\ref{NewDefn}.
  The notion of $\epsilon_{\rho}$-distin\-guishability is defined, and shown  to be
  finer  than  $\epsilon_h$-dis\-tinguishability.  Here $h$ stands for the (partial) Hamming
  metric assumed definable  between the type-compatible tuples concerned.

  %\lft
  $\bullet$  In Section~\ref{Analysis}, which is the core additional contribution to the
  previous version of this article~\cite{siva-etal-2022}, we show that a DLTTS-based
  vision can be   applied  to some practical problems {that a database administrator
  may be confronted with, when analyzing privacy protection using generalization
  in databases},  such as the following:
  
  Assumption: The  `quasi-identifier'  (qid) columns of non-sensitive data in the base
  $D$ are   anonymized via generalization. Adversary $A$ is assumed to have access
  only to the qids,  moreover with specified probability thresholds  on  certain entries.
  \par\indent
  Objective: To estimate the maximal probability threshold for $A$ to get
  access to the sensitive values of some of the entries in $D$ (and possibly taking
  subsequent actions such as restricting access by not answering any more queries). 

  $\bullet$  Related Work and Comments constitute Section~\ref{Conclude}.

  $\bullet$  Section~\ref{Conclusion} concludes the paper.  
     }   %% Fin small

   \textbf{Positioning w.r.t. our previous work.} The work presented
  in the current  paper is an extensively revisited version   of our
  earlier  work~\cite{siva-etal-2022} presented in a conference.
  In that work, the formalism of DLTTS was fully developed, as well as a 
  `value-wise' (partial) metric between  type compatible sets of data,  for  a large
  class of databases. The same role continues to be played by the DLTTS in the
  current paper, with some additional precisions. But the role played by the `value-wise'
  metric  has been {\em significantly modi\-fied} in the current version:  this 
  metric   is  now assumed known  {\em only} to (the system administrator and)   the
  oracle  mechanism $\o$ controlling the runs   on a DLTTS  modeling a query-sequence
  of an adversary on  the base. And, more importantly, contrary to~\cite{siva-etal-2022},
  the oracle $\o$ no longer   gives any  information of any  kind  to the  adversary
  (such as  on how close or how   far (s)he is, from  information  intended to
  remain  secret),  at any node on the   DLTTS modeling  the query  runs of the
  adversary. Thus our new proposition can now be used as an 'assistant' for a
  database administrator to control and limit query execution on a sensitive database.
  In particular, Section~\ref{Analysis} is a new contribution; Section~\ref{DataWise}
  has been extended, and other sections have been restructured and augmented with
  novel DLTTS examples and illustrations to enhance readability.
  Sections~\ref{LDP}, \ref{DP} and \ref{NewDefn} are presented essentially
  as in our previous work~\cite{siva-etal-2022}.

 %\vspace*{-1em}
\section{ {\large Preliminaries}}~\label{prelim}  
%\vspace*{-1.5em}
In this section, we present concepts useful to understand the work presented in
this article, and introduce a simple running example.

\subsection{Useful concepts definitions}
We assume given a database $D$, with its attributes set $\A$, usually divided in 
three disjoint groups: the subgroup $\A^{(i)}$ of {\em identifiers}, $\A^{(qi)}$ of
{\em quasi-identifiers}, and  $\A^{(s)}$ of {\em sensitive attributes}. The tuples of
database $D$ will be generally  denoted as $t$, and their attributes
denoted respectively as $t^{i}, t^{qi},$ and $t^{s}$ in the three subgroups of $\A$.
The attributes $t^{i}$ on any tuple $t$ of $D$  are conveniently viewed  as defining  a
`user' or a `client' stored in database $D$.
Quasi-identifiers\footnote{The notion of quasi-identifier attributes was introduced in
informal terms, by T. Dalenius in~\cite{Quasi-Id}. Suffices, for now, to see them as
attributes that are not identifiers nor  sensitive.} are informally defined as a set of
public attributes, which  in combination with other attributes and/or external
information, can allow to re-identify all or some of the users to whom the information
refers. 

By a  {\em privacy  policy} $P = P_A(D)$ on $D$ with respect to a given
agent/adversary $A$ is meant the stipulation that for a certain {\em given set} of
tuples $\{t \in P \subset D\}$, the sensitive attributes $t^{s}$ on any such  $t$ shall 
remain inaccessible to $A$ (`even after further deduction' -- see below).

The logical framework we propose below, to model the evolution of the `knowledge'
that an adversary $A$ can gain by repeatedly querying the given base $D$ 
will be called  {\em Distributed Labeled-Tagged Transition  System} (DLTTS); repeated 
querying is  intended, in general, to capture sensitive data meant to  remain hidden,
under the privacy policy $P$. The base signature $\Sigma$ for the framework is assumed
to be first-order, with countably many variables, finitely many constants
(including dummy symbols such as `$\star$'), and  no non-constant function symbols.
By `knowledge' of $A$ we shall mean the data that $A$ retrieves as answers to
his/her successive queries, as well as other data that can be derived under
relational operations on these answers, and some others derivable from these using
relational combinations with data (possibly involving certain users of $D$) from
finitely many {\em external  databases  given in advance}, denoted as $B_1, \dots, B_m$,
to which the adversary $A$ has free access. These relational and querying
operations are all assumed done with a well-delimited fragment of the relational 
language SQL; this  {\em SQL fragment is assumed part of the signature  $\Sigma$}. 
In addition, if $n \ge 1$ is the length of the data tuples  in $D$,
finitely many predicate symbols $\K_i, 1 \le i \le n$, each $K_i$ of arity $i$,  will also
be part of the signature $\Sigma$; in the work presented here they will be the only
predicate symbols in $\Sigma$; the role of these symbols is  to allow us to see any data
tuple of length $r, 1 \le r \le n$, as a variable-free first-order formula with top symbol
$\K_r$, with all arguments assumed typed implicitly (with the headers of $D$).
But in practice, we shall drop these top symbols $\K_i$ and see any data tuple
(not part of the given privacy policy $P_A(D)$) directly as a first-order variable-free
formula over $\Sigma$; tuples $t$ that are elements of  $P_A(D)$ are just
written as $\neg t$, in general.  
 We  also assume that  the given external bases $B_1, \dots, B_m$ -- to which $A$
 could resort, for deducing additional information with relational operations -- are 
 of the same  signature $\Sigma$ as $D$. Thus  all the knowledge $A$ can derive 
 from repeated queries on $D$ can be expressed as first-order variable-free
 formulas  over $\Sigma$.

The DLTTS framework will be shown to be well suited for capturing the ideas of acquiring
knowledge and of policy violation, in an elegant and abstract setup.  The definition of this
framework (Section~\ref{DLTTS}) considers only the case where the data, as well as the
answers to the queries, do not involve any notion of `noise'   -- by `noise'  we mean the
perturbation of data by some {\em external}  random mechanism. Sections~\ref{LDP}
and \ref{DP} extend these results to the case where noise can be used as part of the
privacy mechanism.  The DLTTS we consider will be modeling the lookout for the
sensitive attributes of   certain given  users on a base, by a single adversary, with
finitely many queries.
  It is straightforward to extend  the  vision to model query-sequences by
  multiple `non-communicating' users, seeking   to capture possibly different
  privacy policies.  The formal definition of the DLTTS is best motivated by first
  presenting our `running example' in informal style:

\vspace*{-1em} 
\subsection{A Running Example}~\label{RunningExample}

\vspace*{-1.5em} \lft
   {\bf Example 1}. Table~\ref{1} below is the record kept by the central Hospital
   of a  Faculty, on recent consultations by the faculty  staff of three Departments,
   in a University.  `Name' is an identifier  attribute, `Ailment' is  sensitive, the others
   are  QIDs; `Ailment' is categorical with 3 branches: Heart-Disease, Cancer,  and
 Viral-Infection; this latter in turn is categorical too, with 2 branches: Flu and CoVid.
 By convention, such taxonomical relations  are assumed known to public. 
 (For simplicity of the example, we assume that {\em all}  Faculty staff  are on the
 consultation list of the Hospital.) 

 \begin{table}[h]
   \centering
    \begin{tabular}{|c | c| c| c| c|}
  \hline
  Name  & Age  & Gender & Dept.  & Ailment  \\
  \hline
    Joan    &  24  & F  &  Chemistry  & Heart-Disease  \\
    Michel &  46  & M  &  Chemistry & Cancer  \\
    % Aline  &  23  & F  &   Physics     &  Flu    \\
     Aline  &  23  & F  &   Physics     &  Flu  \\
    Harry  &  53  &  M  &  Maths      &  Flu    \\
    John  &  46  &  M  &   Physics    &  CoVid   \\
    \hline
   \end{tabular}
    \caption{\label{1} Hospital's `secret' record }
  \vspace*{-1em}
 \end{table}

 \lft
 The Hospital intends to keep  `secret'  information concerning  CoVid infected
 fa\-culty members; and the tuple {$\neg(John, \star, \star, \star, CoVid)$} 
 is decided  as its privacy policy.   Other privacy policies are of course possible
 (e.g. $\neg(John, 46, M, \star, CoVid)$) and would lead to other analysis formulations.
 Table~\ref{2} is published by the Hospital for the public, where the
 `Age' attributes are anonymized as (integer) intervals
 \footnote{Note that here the generalized interval chosen for Age is non deterministic
 for extremum values.}; and `Ailment' is anonymized  by  an upward push in
 the taxonomy.

 A certain person $A$, who met John at a faculty banquet, suspected John to  have
 been infected with CoVid; (s)he thus decides to consult the published record  of the
 hospital   for  information.  

 \begin{table}[h]
   \centering
    \begin{tabular}{| c| c| c| c|c|}
  \hline
  Line & Age & Gender &  Dept.  & Ailment  \\
  \hline
    $\ell_1$ & $[20-30]$ & F  &  Chemistry  & Heart-Disease  \\
    $\ell_2$ & $[40-50]$  & M  &  Chemistry & Cancer  \\
    $\ell_3$ & $[20-30]$  & F &   Physics     &  Viral-Infection   \\
    $\ell_4$ & $[50-60]$  & M  &   Maths  &  Viral-Infection \\
    $\ell_5$ & $[40-50]$  &  M  &   Physics   &   Viral-Infection   \\
    \hline
   \end{tabular}
    \caption{\label{2} Hospital's published record }
  \vspace*{-1em}
 \end{table}
 
   Knowing that the `John'  (s)he met is `male' and  that Table~\ref{2} must contain
   some information on  John's health, $A$  has as  choice   lines 2, 4 and 5
   ($\ell_2, \ell_4, \ell_5$)   of  Table~\ref{2}.
   $A$ being in the lookout  for a `CoVid-infected  male', this choice is reduced to the last
   two tuples of the table -- which are a priori indistinguishable because of ano\-nymization
   (as `Viral-Infection').  Now, $A$ had the impression  that the John (s)he  met `was not too
   old',  so feels the last tuple $\ell_5$ is twice more likely, and {assumes}    `John must
   be from the   Physics Dept.', so goes to  consult the  following CoVid-cases   record
   `publicly visible' at  the faculty.
   
\begin{table}[h]
   \centering
    \begin{tabular}{| c| c|}
  \hline
  Dept. & CoVid cases  \\
  \hline
      Physics  & M : 1  \, F : 0\\
   \hline
   \end{tabular}
    \caption{\label{3} Faculty CoVid-cases}
   \end{table}
    
 \vspace*{-1em}\lft
And that confirms $A$'s suspicion concerning John.    \hfill $\Box$

\vspace*{1mm}
One of the objectives of this article is to define a formal model to capture this kind of
reasoning. In the next section, we  shall present our Distributed Labeled-Tagged Transition
System (DLTTS) model, which we  believe   is well suited to assist in the modelling and
detection of possible privacy policy breaches; we will show in Section~\ref{Analysis} how
the model can be used by e.g. a database administrator to monitor potential privacy
breaching queries.

 \vspace*{-1em}
 \section{{\large Distributed Labeled-Tagged Transition Systems}}~\label{DLTTS}
 \vspace*{-1.5em}
 
 The DLTTS framework presented in this section synthesizes ideas  coming from 
 several domains, such as the Probabilistic Automata  of Segala~\cite{Segala95b},
 Probabilistic Concurrent  Systems,  and Probabilistic labelled transition systems 
\cite{Fast2018,PTS2019}. Although the  underlying signature for the DLTTS can be
 rich, in the  current paper we shall work with a  limited first-order signature (as
 mentioned in the Introduction) denoted $\Sigma$, with countably many variables,
 finitely many  constants, including certain additional  `dummies', no non-constant
 function symbols,  and  a finite set of propositional (predicate)  symbols.
 Let  $\E$ be the set of all  variable-free formulas  over $\Sigma$, and $\mathtt{Ext}$
 a given subset of $\E$. We assume given a  decidable  procedure $\C$ whose role
 is to `saturate'  any finite set $G$ of  variable-free formulas  into a finite set
 $\ovr{G}$, by adding a finite  (possibly empty) set of variable-free 
formulas, using   {\em  relational  operations} on  $G$ and $\mathtt{Ext}$. This 
procedure $\C$ will be  {\em  internal} at every node on a DLTTS, it is assumed
executed using a given {\em finite set of internal actions}. There will 
be an oracle  $\o$ as mentioned earlier, to `check' if the given privacy policy
on the database is violated at the current node.

In the definition below, $L$ will stand for a given (finite) set of ground (variable-free) 
first-order statements over $\Sigma$, its elements will be called {\em  labels}.
For any set $S$, $Distr(S)$ will stand for the set of all probability  distributions 
with finite support,  over the subsets of $S$. 

\begin{defn}
  A Distributed Labeled-Tagged Transition System (DLTTS), over a given
  signature $\Sigma$, is formed of:
  \begin{itemize}
 \item[-] a  finite (or denumerable) set $S$ of  states, an `initial'  state  $s_0 \in S$,
 and a special  state $\otimes \in S$ named `Stop'. 

\item[-] a finite set $Act$ of action symbols (disjoint from $\Sigma$), with  a
  special action $\delta \in Act$ named `violation'. 
    
\item[-] a (probabilistic) transition relation $\T \subset S\times Act \times Distr(S)$. 

  \item[-] A transition $\t = (s, \alpha, \t(s)) \in \T$ is said to be `from' the state
    $s \in S$,   and every  $s' \in  \t(s)$ is  a $\t$-successor of $s$. The `branch'  of
    $\t$ from $s$ to $s'$ is `labeled' with a label $l(s, s') \in L$. 
  
\item[-] a tag $\tau(s)$ attached to every state  $s \in S$ other than  $\otimes$, 
  formed of  finitely many ground  first-order formulas   over $\Sigma$.
  The tag at  $s_0$  is $\{\top\}$, the tag   at $\otimes$ is the empty set $\emptyset$. 

\item[-] at every state  $s \in S$ other than  $\otimes$  a special action symbol
  $\iota = \iota_s \in  Act$,  {\em internal}  at $s$,  `saturates' $\tau(s)$ into
  a set  $\otau(s)$  using the procedure $\C$. 
 \end{itemize}
\end{defn}

\vspace*{-0.5em}
The formulas in the tag $\otau(s)$ attached to any  state $s$  will all have 
the same probability as assigned (by the distribution) to  $s$. 
 If the set $\otau(s)$ of formulas turns out to be inconsistent,   the oracle
 $\o$ will impose $(s, \delta, \otimes)$ as the only  transition from $s$,
 which stands for `violation'  and `Stop'. No outgoing transition or internal
 action  at  the halting state $\otimes$.

 \vspace*{0.5mm}
{\sc  Remark}~1:
(a) We shall assume our DLTTS to be fully probabilistic, in the following sense:
 Given a state $s$ and a  given distribution  $E' \in Distr(S)$, there is at most
 one probabilistic  transition from $s$ with $E'$ as its set of successors.
 
 (b) It will also be assumed that the {\em tags at the states of a DLTTS are tight, in
   the  following sense}, wrt the labels (and the procedure $\C$): For  any state $s$,
 any  transition $\t$ from $s$ and any  $\t$-successor $s'$ of $s$, we have 
 $\tau(s') = \otau(s) \cup l(s, s')$, except  when $s'$ is the state Stop $\otimes$. 
   
 (c)  One final assumption: `No infinite set can get generated from a finite set'
 by the  procedure $\C$ for  gaining  further knowledge, at any state $s \in S$.
 (This corresponds  to the assumption of {\em bounded inputs outputs}, as  in e.g.,
 \cite{BarthePOPL12,BartheLICS20}.)                  \hfill$\Box$
 
 \vspace*{1mm}
 {\bf DLTTS and Repeated queries on a database}: 
 The states of the DLTTS will stand for the  various  `moments'  of the querying sequence,
 while the  tags attached to the states  will stand for the knowledge $A$ has acquired
 on the  data of $D$ `thus far'.  This knowledge consists partly in the answers to  the
 queries (s)he  made so far,  then saturated  with additional knowledge  using the
 (finitely many) internal relational operations of the procedure $\C$, between the
 answers retrieved (as  tuples/subtuples in $D$) by $A$ for his/her queries,
 and suitable tuples from the given external  databases $B_1, \dots, B_m$.
 If the saturated knowledge of $A$,  namely $\otau(s)$, at a current state $s$ on the
 DLTTS is not inconsistent, then the  transition  from  $s$  to  its successor  states
 represents the probability distribution of the likely  answers  $A$  would  expect
 to get for the next query.
 
 Note that we make no assumption on whether the repeated queries by  $A$ on
 $D$  are treated {\em interactively, or non-interactively}, by the DBMS.  It appears 
 that the logical framework would function exactly alike, in both cases. 

 %%%%%%%%%%%%%%%%%%%
 \lft
  {\bf Example~1}(bis):  Figure~\ref{fig:ex_dltts}  below shows a DLTTS functioning
  on   the   Check-for-CoVid problem of Example~1  above  The edges in the figure are
  marked  with the `query expressions' of $A$, the labels on the branches are part
  of the  answers to $A$'s current queries: 
     
 \vspace*{-1mm}
 \begin{figure}[h]
  \centering
\begin{tikzpicture}[node distance=0.8cm, auto]  
   \node[state,initial, initial text ={}] (s_0)   {$s_0$}; 
   \node[state] (s_1) [above right=of s_0] {$s_1$};
   \node[state] (s_2) [below right=of s_0] {$s_2$}; 

   \node[state] (s_3) [above right=of s_2] {$s_3$};
   \node[state] (s_4) [below right=of s_2] {$s_4$};
  
  \node[state] (s_5) [above right=of s_4] {$s_5$};
  \node[state] (s_6) [below right=of s_4] {$s_6$};
    
  \node[state] (f) [right=of s_6] {$\otimes$};

  \path[->] 
    (s_0) edge  node {$\frac{M : 0} { \{l_1, l_3\} } $} (s_1)
        edge  node [swap] {$\frac{M : 1} { \{l_2, l_4, l_5\} }$} (s_2)
    (s_2) edge node[swap] {$\frac{Covid : 0} {l_2} $} (s_3)
    	edge node[swap]  {$\frac{Covid : 1} { \{l_4, l_5\} }$} (s_4)
    (s_4) edge node[swap]  {$\frac{Not\, Old, \, pr 1/3} {l_4}$} (s_5)
    	edge node[swap]  {$\frac{Not\, Old, \, pr 2/3} {l_5}$} (s_6)	
	(s_6) edge node {$\delta : 1$} (f)    	;
\end{tikzpicture}  

\vspace*{-1.5mm}
\caption{\label{fig:ex_dltts}A DLTTS for Check-for-CoVid}
 \end{figure}
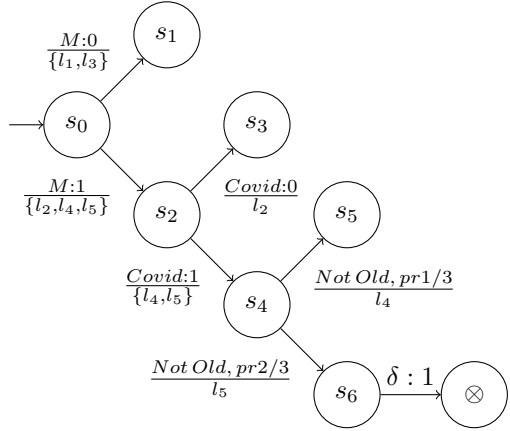
%%%%%%%%%

 \vspace*{-1mm}
  \begin{propn}~\label{exact}
   Suppose given a database $D$, a finite sequence of repeated queries  on $D$ by an 
   adversary $A$, and a first-order relational formula $P = P_A(D)$ over the signature $\Sigma$
   of $D$,  expressing the privacy policy of $D$ with respect to $A$. Let $\W$ be the DLTTS
   modeling the various queries of $A$ on $D$, and the evolution of the knowledge of $A$
   on the data of $D$, resulting from these queries and the internal actions at the states of
   $\W$, as described above.

   (i) The given privacy policy  $P_A(D)$ on $D$ is violated under some run on $\W$, if and
   only if the failure state   $\otimes$ on  $\W$ is reachable on $\W$ under that run. 

      (ii) The satisfiability of the set of formulas $\otau(s) \cup \{\neg P\}$ is decidable,  
   at any state $s$ on the DLTTS, under the assumptions of Remark~1(b). 
  \end{propn}
  
  \vspace*{-0.5mm}\lft
  {\em Proof}: Assertion (i) is just restatement. Observe now,  that at any state $s$ on
  $\W$,  the tags $\tau(s)$, $\otau(s)$ are both finite sets of first-order {\em variable-free
    formulas} over  $\Sigma$, without non-constant function symbols. Indeed,  to start
  with, the knowledge of $A$ consists of the  responses received for  his/her queries,
  in the form of a finite set of data tuples/subtuples  from the  given  bases; and 
  by our assumption of Remark~1(c), no infinite set can be generated by  saturating
  this  initial knowledge  with the procedure $\C$.  
  Assertion (ii) follows then  from the known result that the inconsistency of any given
  finite set of {\em variable-free} first-order Datalog formulas  is  decidable, e.g., by the 
analytic tableaux procedure.   (Only the absence of variables is essential.)  \hfill$\Box$ 

\vspace*{-1em}
 \section{ {\large Value-wise Metrics on Databases} }~\label{DataWise}
 \vspace*{-2em}
 
 We assume given a (distributed) database $D$, with the datatypes as mentioned
 in the Introduction (with a privacy policy $P$ specified on $D$). 
 Our objective in this section is to look for a `quantitative measure' for comparing 
 type comparable tuples in $D$, based on which it might be possible to define a 
 (partial) notion of distance/metric $\rho$ between sets of tuples/subtuples in  $D$. 
 Our objective is motivated by the following considerations. Suppose an adversary
 $A$ launches a sequence of queries on $D$, with a view to capture some of the
 sensitive data in the policy $P$; and suppose a DLTTS models the sequence
 of queries-and-answers (as described in Section~\ref{DLTTS}). Assumed  given
  $\epsilon \ge 0$ `sufficiently small', as  a `threshold' for approximation, the
 manner in which  the DLTTS functions can then be refined  in two  different ways 
as described below:  

 (1) The role of the oracle  mechanism $\o$ can be made `sharper',  than just 
 registering the violation of the policy $P$: if the current (saturated) knowledge
 of $A$ at some node $s$ on the   DLLTS is  at $\rho$-distance $\le  \epsilon$, then
 the  oracle  $\o$ could force the transition  from $s$ to the state $\otimes$ (Stop);
 such a transition would then be named {\em $\epsilon$-violation} of privacy.
 {In operational terms, a DLTTS could inform the database administrator,
   or some access control system, which in turn could take actions such as refusing
   to answer subsequent queries} 
   
 (2)  Two distinct `query instances' by $A$, at some  given node $s$ on the DLTTS, 
 could receive the  same output(answer) from the answering  mechanism, as concerns
 the secret data from $P$. The two  instances will then be said to be
 $\epsilon$-indistiguishable, in a sense  and under certain conditions that  we define
 formally in the  following  section~\ref{LDP}. In such a  case, the labels on the
 outgoing branches at $s$,  corresponding to  the knowledge gained by $A$ from
 these queries, will be  considered  {\em $\epsilon$-equivalent} in $\otau(s)$,  for privacy. 

It will be actually shown in  section~\ref{LDP}, that the notion of $\epsilon$-equivalence
 can be rendered  finer, into an {\em $\epsilon_{\rho}$-equivalence}, by  combining it
 with the value-wise metric $\rho$ constructed  in the current section.

Remember that  the knowledge of $A$, at any node on the DLTTS is 
represented  as a set of tuples, and  that the data forming  any tuple are
assumed `implicitly typed with  the headers' of the  database $D$.
For `quantitatively' comparing two tuples of the same length,  we shall  assume
there is a natural, injective, {\em type-preserving}  map from  one of them onto
the other; this map will  remain  implicit  in general, and two such  tuples will be
said to be {\em type-compatible}. If the two tuples are not of  the same length,
 one of them will be projected onto (or restricted to) a suitable  subtuple, so as to 
 be type-compatible and comparable with the other; if this turns out  to be impossible,
 the two tuples will be said to be uncomparable.

 We shall propose a  comparison method based on an appropriately defined
 notion of `distance'  between  two {\em sets of type-compatible  tuples}.
 For that, we shall first  define  a `distance' between any two type-compatible
 tuples -- more precisely, define a  notion of distance between any two data
 values under every given header of $D$.
 As a first step, we  shall therefore begin by defining, for every given  header  of
 $D$, a binary `distance'  function on the {\em set of all values  that get  assigned
   to the attributes under that header}, all  along the sequence of $A$'s  queries.
 This distance function  to be defined will be a {\em metric}: non-negative,
 symmetric,  and satisfying the  so-called  Triangle  Inequality (cf. below).
 The `{\em direct-sum}' of these metrics, taken over  all the headers of $D$,
 will then define a metric $d$ on the set of all  type-compatible  tuples  of data
 assigned to the various attributes, under all the headers of $D$, all along the
 sequence of  $A$'s queries.
 The `distance'  $d(t, t')$, from any given tuple $t$  in this set to another
 type-compatible  tuple $t'$,  will be  defined as the value of  this direct-sum 
 metric on the pair of tuples $(t, t')$; it will be calculated  `column-wise' by 
 definition, on $D$ and also on the intermediary databases along  $A$'s
 query sequence. Note that it will a priori give us an $m$-tuple of numbers,
 where  $m$ is the  number  of headers (number of columns) in the
 database $D$.

 A single number can then be derived  as the  sum of the  entries in  this
 $m$-tuple $d(t, t')$. This sum will be denoted as $\ovr{d}(t, t')$, and defined
 as the distance from the tuple $t$ to the tuple $t'$ in the database $D$. 
 Finally, if $S, S'$ are any two given finite sets of type-compatible tuples, of data
 that  get assigned to the various attributes (along the sequence of $A$'s queries),
 we define  the distance from the set $S$ to the set $S'$ as the number 
  $\rho(S, S') = min \{\, \ovr{d}(t, t') \mid t \in S, \, t' \in S' \, \}$

 \vspace*{1mm}
 Now, for clarity of presentation, in order to define the `distance' between the  data
 values under  every given header of $D$, we  divide the  headers of $D$ into  four
 classes, as below:  
   
 \vspace*{-1em}
 \begin{itemize}
 \item[.]  `Nominal':  identities, names, attributes  receiving {\em literal} data 
  {\em not  in any taxonomy} (e.g., gender, city, \dots), and finite sets  of  such data;
   \vspace*{-1mm}
 \item[.]  `Numerval' : attributes receiving {\em numerical} values, or bounded
   intervals of  (finitely many) numerical values;
  \item[.]   `Numerical': attributes receiving {\em single numerical values}  (numbers).  
   \vspace*{-1mm}
 \item[.] `Taxoral': attributes receiving {\em literal data in a taxonomy relation}. 
\end{itemize}

 \vspace*{-1mm}
 $\bullet$  For defining the `distance'  between any two values $v, v'$ assigned to an
 attribute under a given `Nominal'   header of $D$,  for the  sake of uniformity  we
 agree to  consider every value as a  {\em  finite set}  of singleton values; in particular,
 a  singleton value `$x$' will be seen as the set $\{x\}$. Given two such values
 $v, v'$, note  first that the so-called {\em Jaccard Index} between them is the number
 $jacc(v, v') = |(v \cap v') / (v \cup v') |$, often called  a `measure of their similarity';
 but this index  is not a metric, because  the {\em triangle inequality} is not satisfied;
 however, the  Jaccard metric $d_{Nom}(v, v') = 1 - jacc(v, v') =  |(v \Delta v') / (v \cup v')|$
 does satisfy that property, and will suite our purposes. Thus defined, $d_{Nom}(v, v')$
 is a `measure of the dissimilarity' between the sets $v$ and $v'$. 
   
 $\bullet$ Let  $\Tau_{Nom}$  be the  set of all data assigned to the attributes under
 the  `Nominal' headers of $D$, along  $A$'s queries sequence. Then the
 above  defined binary  function $d_{Nom}$ extends to a metric on the set   of all
  type-compatible data-tuples from $\Tau_{Nom}$, defined as the `direct-sum' 
  taken over the `Nominal' headers of $D$. 
    
  $\bullet$ If $\Tau_{Num}$  is the set of all data assigned to the attributes under  the
  `Numerval'   headers along the sequence  of queries by $A$, we  define in a similar
  manner (as above)  a `distance'  metric  $d_{Num}$ on the set of all  type-compatible
  data-tuples  from $\Tau_{Num}$:  we first define $d_{Num}$ on any couple  of values
  $u, v$  assigned to  the attributes under a given `Numerval' header of $D$, then
  extend it  to the set of all type-compatible data-tuples from $\Tau_{Num}$ (as the
  direct-sum  taken  over the  `Numerval'   headers of $D$).  This will be done exactly
  as  above under  the `Nominal' headers: suffices to visualize any finite interval value
  as a  particular   way of presenting a set of numerical values (integers, usually).
  In particular, a   single value `$a$'    under a `Numerval'  header will be seen as
  the interval value   $[a]$.) Thus defined the (Jaccard) metric $d_{Nom}([a, b], [c, d])$
  will be a   measure of `dissimilarity'  between $[a, b]$  and $[c, d]$. 

  $\bullet$ Between numerical data $x, x'$ under the `Numerical' headers, the distance
  we shall    work with is the euclidean  metric $|x - x'|$,  {\em normalized as}: 
   $d_{eucl}(x, x') = |x - x'| / D$, where  $D > 0$ is a fixed finite number,  bigger than
   the maximal euclidean  distance between the numerical data on the databases 
   and on the answers to $A$'s queries. 

   $\bullet$ For the data under the `Taxoral'  headers, we choose as distance function the 
   metric    $d_{wp}$ that we define  in the Appendix  (Section~\ref{WP-Metric}),  based on 
   the  well-known  notion of Wu-Palmer symmetry between the nodes of a Taxonomy tree,
   
   \vspace*{1mm}
    {\sc  Remark}~2: 
   (a) Note that  the `datawise  distance functions' defined above  are all {\em with values
    in   the real interval} $[0, 1]$.  This is one reason for our choice of the distance 
   metric on Taxonomy trees, it is of importance, cf. Section~ \ref{NewDefn}.
   
   (b)  The Hamming metric between datatuples (and databases) is generally well-defined
   only for   databases with all data of a single (numerical or string) type, and all tuples
   of the   same length.  However, in   this paper we shall be using  a generalized notion
   of that   metric, by extending that  usual notion, in a natural manner `data-wise' and
   `column-wise' as a partial metric,  just as we did for the distance function $\rho$
   above. We shall denote it  as $d_h$. For instance, we shall have
   $d_h([1,2], a), ([2,3], a))=1, d_h([1,2], a), ([2,3], b))=2$, whereas
   $d_h((bd, a), ([2,3], b))$ is  undefined, etc. 
   
 \vspace*{-1.2em}
\section{{\large   $\epsilon$-local-differential  privacy,
      $\epsilon$-indistinguishability}}~\label{LDP}
 \vspace*{-1.5em}
 
 In this section we extend the result of  Proposition~\ref{exact} to cases  where  the
 violation of a policy  can be  up to  a `threshold of approximation' $\epsilon \ge0$,
 in the sense  defined in the previous section.
 We stick to the same notation as above. The  set  $\E$  of all variable-free formulas
 over $\Sigma$ is thus a disjoint union of  subsets  of the  form
 $\E = \cup \{\E^{\K}_i \mid 0 < i \le n, \K \in \Sigma\}$, the  index  $i$ in $\E^{\K}_i$
 standing for  the common length of the formulas in the subset, and $\K$  for the
 common  root symbol of its formulas; each  set $\E^{\K}_i$  will be seen as a
 database  of $i$-tuples. 
  
 As above, we consider the situation where the queries of an adversary intend  to capture
 certain  given (sensitive)  values  in the database $D$. The following  definitions of
 $\epsilon$-indistinguishability (and of $\epsilon$-distinguishability) of two different
 query instances for the answering mechanism $\M$,  as well as that of $\epsilon$-DP
 that will be defined in the next subsection, are essentially reformulations of the
 same (or similar)  notions defined in  \cite{Dwork2006,Dwork2014}.
 
  \vspace*{-1mm}
 \begin{defn}~\label{eps-indistinguish}
   (i) Suppose the probabilistic mechanism $\M$,  answering $A$'s queries  on the
   base $D$   outputs(answers with) the same  tuple  $\alpha \in \E$ for  two different
   input   instances $v, v'$.   Given $\epsilon \ge 0$, the two instances will be said
   to be  $\epsilon$-indistinguishable wrt $\alpha$, if and only if: \par   
   \hspace*{2.5cm} $ Prob[\M(v) = \alpha]  \le e^{\epsilon} Prob[\M(v') = \alpha]$ and  \par 
   \hspace*{2.5cm} $ Prob[\M(v') = \alpha]  \le e^{\epsilon} Prob[\M(v) = \alpha]$. \par\lft 
   Otherwise, the two  instances  $v, v'$ are said to be $\epsilon$-distinguishable
   for output $\alpha$. 
   
   (ii) The probabilistic answering mechanism $\M$ is said to  satisfy
   {\em $\epsilon$-local differential privacy} {\em($\epsilon$-LDP)}  for $\epsilon \ge 0$,
   if and only if: for any two instances  $v, v'$ of   $\M$ {\em that lead to  the same
     output},   and any set   $\S \subset Range(\M)$, we have \par 
   \hspace*{2.5cm} $ Prob[\M(v) \in \S ]  \le e^{\epsilon} Prob[\M(v') \in \S]$.
   \end{defn}

   \vspace*{-1.5mm}
  The two small examples below illustrate  $\epsilon$-lndistinguishability:

  (i) The  two queries based on sub-tuples ([50--60], M, Maths) and   ([40--50], M,
  Physics), from the Hospital's  published record in Example~1  (Table~\ref{2}) 
  have both Viral--Infection as output, with respective  probabilities  $1/3, 2/3$;
  thus they are  $\epsilon$-indistinguishable for any  $\epsilon \ge ln(2)$; and
  $\epsilon$-distinguishable for any $0 \le \epsilon < ln(2)$.

   (ii) The `Randomized Response' mechanism $RR$~(\cite{Warner65}) can be 
  modelled as follows. Input is ($X, F_1, F_2$) where $X$ is a Boolean, and $F_1, F_2$
  are flips of a coin ($H$ or $T$). $RR$ outputs $X$ if $F_1=H$, $True$ if
  $F_1=T$ and $F_2=H$, and $False$ if $F_1=T$ and $F_2=T$. This mechanism is
  $ln(3)$-LDP : the instances ($True, H, H$), ($True, H, T$), ($True, T,
  H$) and ($True, T, T$) are $ln(3)$-indistinguishable for output $True$.
  ($False, H, H$), ($False, H, T$), ($False, T, H$)and ($False, T, T$) are
  $ln(3)$-indistinguishable for output $False$.

 \vspace*{-1em}
 \section{ {\large $\epsilon$-Differential Privacy} }~\label{DP}
  \vspace*{-2em}

  The  notion of  {\em $\epsilon$-indistinguishability  of two given databases} $D, D'$ 
  for an  adversary, is more general than that of   $\epsilon$-indistinguishability
  of pairs of query instances  giving the same output (defined above).
  $\epsilon$-indistinguishability for  pairs of databases $D, D'$ is usually  defined
  only for bases that are {\em adjacent} in a certain sense (cf. below).  

  There seems to be no uniquely defined  notion of adjacence on pairs of databases;
  in fact,   several  are  known and  in use in the literature. Actually, a notion of
  adjacence can  be   defined  in a generic  parametrizable  manner (as in
  e.g., \cite{dpMetrics2013}),  as follows.
  Assume given a map $\brho$ from the set $\D$ of all databases of 
  $m$-tuples  (for some given $m > 0$), into some given metric space $(X, d_X)$.
  The (symmetric) binary  relation  on pairs of databases in $\D$,  defined by 
  $\brho_{adj}(D, D') =  d_X(\brho(D), \brho(D'))$   is then said to give a {\em
    measure of adjacence}  between these  bases. The relation $\brho_{adj}$ is
  said  to define an `adjacency relation'. 
   
  \begin{defn} %~\label{eps-indistinguish2}
    Let  $\brho_{adj}$ be a given  adjacency  relation on a set $\D$ of databases, and 
    $\M$ a probabilistic answering mechanism for queries on the bases  in $\D$.
    Two bases $D, D' \in \D$ are said  to be   $\brho_{adj}$-indistinguishable
    under $\M$,  if and only if, for any   possible output  $\S \subset Range(\M)$,  we
    have
    
     \disp{$ Prob[\M(D) \in \S ]  \le e^{\brho_{adj}(D, D')} Prob[\M(D') \in \S]$ \par 
                    $ Prob[\M(D') \in \S ]  \le e^{\brho_{adj}(D, D')} Prob[\M(D \in \S]$.}
     
     \lft The mechanism $\M$ is said to  satisfy     {\em $\brho_{adj}$-differential privacy}
     ($\brho_{adj}$-DP),  if and  only if the above conditions are   satisfied for  {\em every
     pair of  databases}   $D, D'$ in $\D$, and  any possible output  $\S \subset  Range(\M)$. 
    \end{defn}
      
    \lft 
    {\em Comments}: (i) Given $\epsilon \ge 0$, the `usual' notions of {\em
      $\epsilon$-indistinguishability and $\epsilon$-DP} correspond, in general, to the
    choice of  adjacency $\brho_{adj} = \epsilon d_h$,  where $d_h$ is the
    {\em Hamming metric}  on databases (namely, the number of `records'
    where $D$ and $D'$   differ) and the assumption $d_h(D, D') \le 1$,
    (cf. \cite{dpMetrics2013}).  But in this paper we shall be using a  generalized
    version of that  Hamming metric, well-defined as a (partial) metric between
    all bases concerned, as e.g. between the tuples and subtuples  of the
    database of our running Example~1.) 

    (ii) In Section~\ref{NewDefn}, we propose a more general notion of adjacency 
    based on the value-wise (data-wise) metric $\rho$ we defined in Section~\ref{DataWise}. 
    
  (iii) On disjoint databases, one can work  with  different adjacency relations,  using
  different maps to the same (or different) metric  space(s), 

  (iv) The mechanism $RR$  described above is actually $ln(3)$-DP,  not only
  $ln(3)$-LDP. To check $DP$,   we have to
  check   all possible  pairs of numbers of the form  $(Prob[\M(x) = y], Prob[\M(x') = y])$, 
  $(Prob[\M(x) = y'], Prob[\M(x') = y])$,  $(Prob[\M(x) = y], Prob[\M(x') = y'])$, etc.,
  where   the $x, x'. ...$ are the input instances  for  $RR$, and   $y, y',
  ...$ the outputs.  The mechanism $RR$ has $2^3$ possible input instances for
  ($X, F_1, F_2$) and two  outputs ({\em  True, False}); we thus have 16   pairs of
  numbers, among which the distinct ones are: $(1/4, 1/4), (1/4, 3/4),  (3/4,  1/4)$, 
  and $(3/4, 3/4)$;  if  $(a, b)$ is   any such pair, obviously $a  \le  e^{ln(3)} b$.
  Thus $RR$ is indeed $ln(3)$-DP.    \hfill$\Box$

\vspace*{-1em}
 \section{{\large Metrics for Indistinguishability and DP}}~\label{NewDefn}
\vspace*{-1.5em}

Given a probabilistic mechanism $\M$ answering the queries on  databases,
and an $\epsilon \ge 0$, recall that the   $\epsilon$-indistinguishability  of any two given
databases under $\M$, and  the notion  of $\epsilon$-DP for  $\M$,  were both defined 
in Definition~\ref{eps-indistinguish} (Section~\ref{DP}); based first on  a
 hypothetical map $\brho$ from the set of all the databases concerned, into some
  given  metric space $(X, d_X)$, and an `adjacency relation' on databases defined as
$ \brho_{adj}(D, D') = d_X(\brho D, \brho D')$,  which was  subsequently instantiated to
  $\brho_{adj} = \epsilon d_h$, where $d_h$ is the (generalized) Hamming metric between
  the bases concerned,  cf. Remark~2(b)

  In this section, our objective is to propose a more general notion of adjacency,
  based  on  the  metric $\rho$ we defined in Section~\ref{DataWise}, between
  type-compatible tuples on databases   with   data of multiple types.
  In other words, our $\D$ here will be  the set of all   databases, {\em not necessarily
    all with the same number of columns, and also with data of   several possible types}
  as mentioned in the Introduction. We define then  a (partial) binary   relation
  $\brho^{\rho}_{adj}(D, D')$ between databases $D, D'$ in the set  $\D$ by 
  setting $ \brho^{\rho}_{adj}(D, D') = \rho(D, D')$, visualizing  $D, D'$  as  sets of
  type-compatible data tuples. 
 
  Given $\epsilon$, we can then define the notion of $\epsilon_{\rho}$-indistinguishability
  of   two  databases $D, D'$ under a (probabilistic) answering mechanism $\M$, as well as
  the  notion of $\epsilon_{\rho}$-DP for $\M$, exactly as in
  Definition~\ref{eps-indistinguish},
  by   replacing   $\brho_{adj}$ first with the relation $ \brho^{\rho}_{adj}$,  and
  subsequently with   $\epsilon\rho$. The notions thus defined are {\em more general} than 
  those presented   earlier in  Section~\ref{DP} with the choice  $\brho_{adj} = \epsilon  d_h$.
 An example will illustrate this point.  
  
  \vspace*{1mm}\lft
  {\bf Example 5}. We go back  to  the `Hospital's public record' of our  previous
  Example~1, and the two sub-tuples ([50--60], M, Maths) and   ([40--50], M, Physics),
  from  the Hospital's  published record in Example~1  (Table~\ref{2}). The  mechanism
  $\M$ answering two queries for `Virus-Ailment   information involving men', returns
  the tuples  $l_4, l_5$ with the  probability distribution $1/3, 2/3$, respectively.
Let us  look for the minimum value of $\epsilon \ge 0$, for which these tuples
 will be $\epsilon_{\rho}$-indistinguishable under the mechanism $\M$.

 We first compute the  $\rho$-distance between the two tuples: \\ \hspace*{2cm}
 $\rho(l_4, l_5) =  \ovr{d}(l_4, l_5) = (1 - \frac{1}{20}) + 0 + 1 + 0 = 39/20$.
 
  The condition   for $l_4$ and  $l_5$ to be $\epsilon_{\rho}$-indistinguishable
  under $\M$ is thus: \\ \hspace*{2cm}
    $ (1/3 ) \le e^{(39/20) \epsilon}*(2/3),  \;\; \;\;  (2/3) \le  e^{(39/20)\epsilon}*(1/3)$. 

  \lft
  Which gives: $\epsilon \ge (20/39)*ln(2)$. That is, for $\epsilon \ge (20/39)*ln(2)$, the
  two tuples $l_4$ and $l_5$ will be $\epsilon_{\rho}$-indistinguishable; and for  values of
  $\epsilon$ with $0 \le \epsilon <   (20/39)*ln(2)$, these tuples will be
  $\epsilon_{\rho}$-distinguishable.

  Now, the Hamming metric is definable between these two tuples: they differ
  only  at two places, we have $d_h(l_4, l_5) =  2$.  So, they are 
  $\epsilon_h$-indistinguishabilty (wrt $d_h$),  for $\epsilon  \ge 0$,  if and only if: 
  $(2/3)  \le  e^{2 \epsilon}*(1/3)$,\, i.e., \, $\epsilon \ge (1/2)*ln(2)$ .

 In other words, if these two tuples are $\epsilon_{\rho}$-indistinguishable  wrt $\rho$
 under $\M$ for some $\epsilon$, then they will  be $\epsilon_h$-indistinguishable wrt
 $d_h$ for  the same  $\epsilon$. But the converse is not true,  since  
 $(1/2)*ln(2) <  (20/39)*ln(2)$. Said otherwise: {\em  $\M$   $\epsilon$-distinguishes
  more finely when combined with  $\rho$, than with $d_h$}.  \hfill$\Box$

 \vspace*{1mm}
  {\sc Remark}~3:  The statement ``$\M$  $\epsilon$-distinguishes  more finely with 
  $\rho$, than with $d_h$'',  is {\em always true} -- not just in Example~4 -- For the
  following  reasons: the records  that differ `at some  given position' on  two bases
  $D, D'$  are  always at distance $1$ for  the Hamming  metric $d_h$, by definition,
  whatever be  the  type of data stored at that position.
  Now, if the  data stored  at that position `happened to be'  numerical, the usual 
  euclidean distance  between the  two data could have been (much) bigger  than
  their Hamming distance $1$; precisely to avoid such  a situation, our definition  of
  the metric $d_{eucl}$ on numerical data `normalized'  the euclidean  distance,
  to ensure that their $d_{eucl}$-distance will not exceed their Hamming distance.
  Thus, all the `record-wise' metrics  we have defined above  have their values in
  $[0,1]$,  as we mentioned earlier; so, whatever the type of data at corresponding
  positions on any two bases $D, D'$, the $\rho$-distance between the records will
  never exceed their Hamming distance. That suffices to prove our  statement above. 
 The Proposition below formulates all this, more precisely:

  \begin{propn}
    Let  $\D_m$ be the set of all databases  with the same number $m$ of columns,
    over a finite set of given data, and $\M$ a probabilistic mechanism answering
    queries on the  bases  in $\D$.  Let $\rho$ be the metric (defined above) and  $d_h$
    the Hamming metric, between the databases in $\D$, and suppose  given
    an $\epsilon \ge 0$. 
    
    - If two databases $D, D' \in \D_m$ are $\epsilon_{\rho}$-indistinguishable under $\M$
    wrt $\rho$, then they are also  $\epsilon$-indistinguishable under $\M$ wrt $d_h$.
   
    - If the mechanism $\M$  is  $\epsilon_{\rho}$-DP on the bases in $\D_m$ (wrt $\rho$),
    then it is also $\epsilon$-DP (wrt $d_h$) on these bases.   
  \end{propn}

  \vspace*{-1mm}
  The idea of `normalizing'  the Hamming metric between numerical databases (with
  the  same number of columns) has already been suggested in several works (cf. e.g., 
  \cite{dpMetrics2013}) for the same   reasons. When only numerical databases are
  considered, the metric   $\rho$ that   we have defined above  is  the same as the
  `normalized Hamming   metric' of  \cite{dpMetrics2013}.  Our metric  $\rho$ must
  actually be seen as a generalization   of that  notion, to directly handle  bases
  with more general  types of  data,  such as anonymized,  taxonomies, etc.  

  \vspace*{-1em}
 \section{Using DLTTS for privacy analysis}~\label{Analysis}
 \vspace*{-1em}

 Our aim now  is to show that a DLTTS-based  vision can be applied to certain practical
 problems, such as the following: \par 
 - Given: a database $D$  with its  qid columns   anonymized. And
   an  adversary $A$ with access only to (some among) the qids, and  specified
   probability thresholds on the entries in $D$.  \par 
   -  Objective: To estimate the maximal probability threshold for $A$'s getting access
      to the sensitive values of (some of) the entries in $D$.

The ideas needed for achieving this objective are all best brought out by the
simple and concrete  example presented  next: 

\subsection{A Motivating example}
\lft {\sc Example}:
 An enterprise $\mathbf{E}$ stores a  database $D$, containing a  sensitive value as
 an integer  between 1 and 10, standing for the (anonymized)  responses of its  employees
 to a  questionnaire on their working conditions.  The qid--attributes of $D$ are  {\em Id,
   Sex, Age}, the anonymized  sensitive value  is `Response'. 
 {\em Age} and {\em Id}  may be anonymized when $D$ is rendered public. 

%\vspace*{-1mm}
\begin{center}
\begin{tabular}{|c|c|c|c|}
\hline
\textbf{id} & \textbf{Sex} & \textbf{Age} & \textbf{Response} \\
\hline
$\ell_1$ & F & [30, 40] & 1\\
\hline
$\ell_2$ & F & [30, 40] & 8\\
\hline
$\ell_3$ & M & [30, 40] & 3\\
\hline
$\ell_4$ & M & [40, 50] & 7\\
\hline
\end{tabular}
\end{center}

\paragraph{Objective:}  {\em An estimation for the maximal probability, for different
  attackers to infer the response  of  one or more randomly chosen employees of
  $\mathbf{E}$, with runs on suitably constructed DLTTS}.
  
With  such a  purpose, the administrator of $D$ conceives three different `test attacks',
where the initial  knowledges/beliefs  of the attackers  on  the qid's are specified, so
as to be `sufficiently complementary'. Based on these tests, an empirical  strategy will
be formulated  in subsection~\ref{Strategy}, with  some  parametric conditions for
accepting/refusing a given  query for accessing any {\em given} response. 
%We present below the analysis  of attacks by  three such different test attackers $A, B, C$.
It is assumed that all the employees in $\mathbf{E}$ have responded to the questionnaire;
none of the attackers  $A, B, C$ is assumed to have any a priori knowledge of the
`Response' of any particular employee.  

\subsection{Privacy attack analysis}
We detail next how DLTTS can be used to analyze privacy attack: we present
several realistic attackers, and compare them.

 \lft\textbf{Evaluating the success of an attack:}
  There are two types of attackers~: Attackers that have some knowledge of
  the individual in the database for whom they wish to retrieve information (Attackers $A$
  and $B$ below), and Attackers with no knowledge other than the database distribution
  (Attacker $C$). The efficiency of an attack is evaluated by first computing (using the
  DLTTS) the probability that an attacker interested in re-identifying a given individual
  will have, using no external knowledge. This gives a baseline probability of success.
  We then compute (using a different DLTTS) the maximal probability with which
  an `informed'  attacker  retrieves this knowledge. If such an attack has a `better'
  probability  of retrieving the correct knowledge, then that attack is considered
  as a success.   (This notion will be formalized in Section~\ref{Strategy}.)

\vspace*{-1.3em}
\paragraph{Attacker $A$} works with the following initial knowledge(belief): 
 \par  \hspace*{6mm} Sex = F : $80\%$,        \hspace*{2cm}  Age = [30-40] : $70\%$ 
 \par  \hspace*{6mm} Sex = M : $20\%$         \hspace*{2cm}  Age = [40-50] : $30\%$
 
  \lft
$\bullet$ $A$ aims to capture `preferably' the response of a female employee,
age 30 to 40 years.

\vspace*{-1.5em}
\paragraph{Attacker $B$} works with the following  initial knowledge(belief): 
\par  \hspace*{6mm} Sex = F : $20\%$      \hspace*{2cm}  Age = [30-40] : $75\%$ 
\par  \hspace*{6mm} Sex = M : $80\%$     \hspace*{2cm}  Age = [40-50] : $25\%$ 

\lft
$\bullet$ $B$ aims to capture `preferably' the response of  a male employee,  age 30
to 40 years.

\vspace*{-1.5em}
\paragraph{Attacker $C$} works with Initial qid knowledge inferred from the base  : 
\par  \hspace*{6mm} Sex = F : $50\%$    \hspace*{2cm} ~Age = [30-40] : $75\%$
\par  \hspace*{6mm} Sex = M : $50\%$   \hspace*{2cm} Age = [40-50] : $25\%$ 

\lft
$\bullet$ $C$ to capture the response of  `any' employee of `any' sex, and of
`any' age. {$C$ has no knowledge beyond the distribution of the qid values in the
  database. In particular, $C$ has no knowledge about possible correlations between
  the QIDs.}\\ \hspace*{1cm}
(We shall  agree to rename attacker $C$ the ``{\em Basic Analyser}''). 

\vspace*{-1mm}
\paragraph{Preliminary Remarks for attack analysis}
For a DLTTS-analysis on privacy policies with probability thresholds  as intended above,
it is natural (even necessary) to `compare' the probability distributions available
prior to choosing a particular transition. Given an outgoing transition $\tau$ at any
given node $s$ on a DLTTS, its distribution is a {\em multiset} of the probability
measures on the branches of $\tau$ that we denote as $\M_{\tau}$. Since our DLTTS
are assumed to be `fully probabilistic', the transition $\tau$ is uniquely determined
by $\M_{\tau}$ and the set $\tau(s)$ of successor states to $s$ (and the labels on the
branches from $s$ to its successors).

\vspace*{1mm}
Now, the multisets of probability distributions (for the transitions available  at any given
node on a DLTTS) are totally ordered by the multiset extension $\succ$ of the natural
order $>$ on  numbers (reals or integers). If $\tau, \tau'$ are two outgoing transitions
from a state $s$ on a DLTTS, we shall {\em define $\tau$ to have
  priority over $\tau'$} if and only if   $\M_{\tau} \succ \M_{\tau'}$.
Two different outgoing transitions from a state $s$ can have  the same multiset
distribution, but with different successor states; if so,  neither has priority
over the other, they will not be `$\succeq$-distinguishable'.

For computing the maximal probability threshold for any attack to capture sensitive
data, it is thus necessary to choose, at any given node on a DLTTS, an outgoing transition
$\tau$ such that $\M_{\tau}$ is maximal for the ordering $\succ$. This will be the
case in the attacks presented below.

On the other hand, `tags'  do not play any role  in the example considered in this
subsection, so the DLTTS constructed for this example make no references to tags
and saturation with `outside' knowledge). 

\vspace*{0.3mm}
On any DLTTS $\A$ constructed below, and any state $s$ on $\A$ where the 
{\em incoming transition has a singleton label set} of the form $\{ \ell \}$, a special
outgoing transition,  referred to as  $response(\ell)$, is assumed  available; its objective
is to give access to the `response'  for the entry $\ell$ in base $D$. This special outgoing
transition $response()$  is in fact a switch, `turned ON, by default'.  (But it can be turned
OFF, at certain states  by the oracle mechanism $\o$ of the DLTTS,  on `considerations of 
strategy for secrecy' -- that we shall present/discuss in section~\ref{Strategy} below.
Pending that discussion, the switch  response() is assumed ON by default; it will be 
an outgoing transition with probability $1$.)

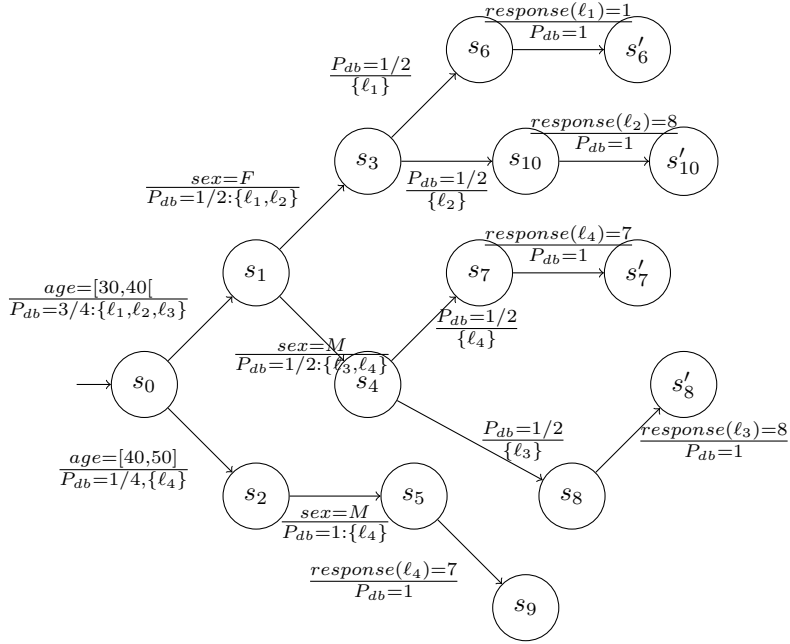
\begin{figure}%[h]
\begin{tikzpicture}[node distance=2.1cm,on grid,auto] 
   \node[state,initial, initial text ={}] (s_0)   {$s_0$}; 
   \node[state] (s_1) [above right=of s_0] {$s_1$};
   \node[state] (s_2) [below right=of s_0] {$s_2$}; 
   \node[state] (s_3) [above right=of s_1] {$s_3$}; 
   \node[state] (s_4) [below right=of s_1] {$s_4$}; 
   \node[state] (s_5) [right=of s_2] {$s_5$};
   \node[state] (s_6) [above right=of s_3] {$s_6$};
   \node[state] (s'_6) [right=of s_6] {$s'_6$};
     \node[state] (s_{10}) [right=of s_3] {$s_{10}$};
     \node[state] (s'_{10}) [right=of s_{10}] {$s'_{10}$};
   \node[state] (s_7) [above right=of s_4] {$s_7$};
   \node[state] (s'_7) [right=of s_7] {$s'_7$};
   \node[state] (s_8) [right=of s_5] {$s_8$};
   \node[state] (s'_8) [above right=of s_8] {$s'_8$};
   \node[state] (s_9) [below right=of s_5] {$s_9$};
    \path[->] 
    (s_0) edge  node {$\frac{age = [30, 40[}{P_{db}=3/4 : \{\ell_1, \ell_2, \ell_3\}}$} (s_1) 
    (s_0) edge  node  [below left] {$\frac{age = [40, 50]}{P_{db}=1/4, {\{\ell_4\}}}$} (s_2)
    (s_1)  edge node {$\frac{sex = F}{P_{db}=1/2 : \{\ell_1, \ell_2\}}$} (s_3)
    %    (s_1) edge node [right] {$\frac{sex = M}{P_{db}=2/3 : \{\ell_2, \ell_3\}}$} (s_4)
    (s_1) edge node [below] {$\frac{sex = M}{P_{db}=1/2 : \{\ell_3, \ell_4\}}$} (s_4)
     %  (s_3) edge node [above] {$\frac{response = 1}{P_{db}=1/2 : \{\ell_1\}}$} (s_6)
          (s_3) edge node {$\frac{P_{db}=1/2}{\{\ell_1\}}$} (s_6)
          %    (s_6) edge node [right] {$\frac{response(l_1}= 1}{P_{db}=1: \{\ell_1\}}$} (s'_6)
          (s_6) edge node [above]{$\frac{response(\ell_1)= 1}{P_{db}=1}$} (s'_6)
          (s_3) edge node [below] {$\frac{P_{db}=1/2}{\{\ell_2\}}$} (s_{10})
           (s_{10}) edge node [above] {$\frac{response(\ell_2)=8}{P_{db}=1}$} (s'_{10})
    (s_2) edge node [below] {$\frac{sex = M}{P_{db}=1 : \{\ell_4\}}$} (s_5)
      (s_4) edge node [right] {$\frac{P_{db}=1/2}{\{\ell_4\}}$} (s_7)
        (s_7) edge node [above] {$\frac{response(\ell_4)=7}{P_{db}=1}$} (s'_7)
     (s_4) edge node [right] {$\frac{P_{db}=1/2}{\{\ell_3\}}$} (s_8)
       (s_8) edge node [right] {$\frac{response(\ell_3)=8}{P_{db}=1}$} (s'_8)
    (s_5) edge node [below left] {$\frac{response(\ell_4) = 7}{P_{db}=1}$} (s_9)
    ;
\end{tikzpicture}
\caption{\label{f1}DLTTS-C: Basic Analyser $C$ captures all responses}
\vspace*{-1em}
\end{figure}

\vspace*{-1.4em}\lft
\paragraph{Basic Analyser $C$:} We first  consider  the case of Basic Analyser
$C$, with initial knowledge inferred from the base $D$, and its QIDs.
The DLTTS-C (Figure~\ref{f1}) shows how  $C$ gets access  to the responses
of all  employees, and respective probability thresholds. The DLTTS construction is
based only on the database $D$; the probabilities on its various branches, denoted as
$P_{db}$,  are all inferred  from $D$.

Starting from the initial state $s_0$ on  DLTTS-C, the probability for access to the
response of employee $\ell_1$ is the product of the probabilities along the branches
traversed by the run, namely: $(3/4)*(1/2)*(1/2) = (3/16)$,which is also  same for
access to the response for entry $\ell_2$;  access to the responses for $\ell_3$ and
$\ell_4$ are  also the same. 
The transitions of DLTTS-C are of $\succ$-maximal priority at all the nodes.
The maximal probability threshold  for access to {\em any of the four responses}, as
reported by Basic Analyser $C$, turns out to be $3/16$. 

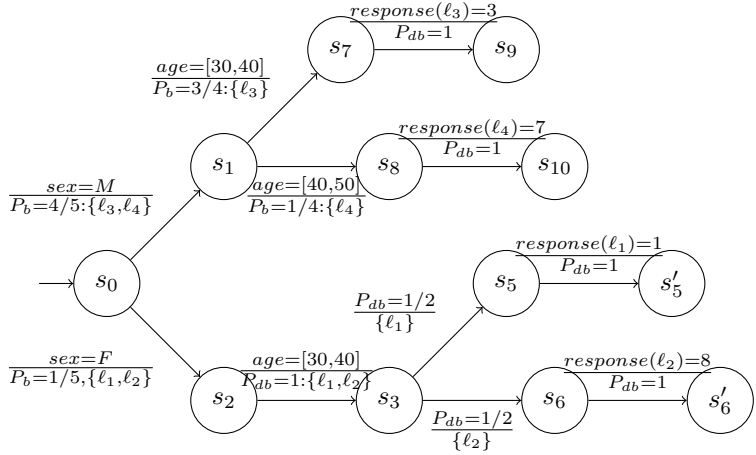
\begin{figure}%[t]
\begin{tikzpicture}[node distance=2.2cm,on grid,auto] 
   \node[state,initial, initial text ={}] (s_0)   {$s_0$}; 
   \node[state] (s_1) [above right=of s_0] {$s_1$};
   \node[state] (s_2) [below right=of s_0] {$s_2$}; 
   \node[state] (s_3) [right=of s_2] {$s_3$}; 
   %  \node[state] (s_4) [right=of s_1] {$s_4$}; 
   \node[state] (s_5) [above right=of s_3] {$s_5$};
   \node[state] (s'_5) [right=of s_5] {$s'_5$};
   \node[state] (s_6) [right=of s_3] {$s_6$};
   \node[state] (s'_6) [right=of s_6] {$s'_6$};
  \node[state] (s_8) [right=of s_1] {$s_8$};
   \node[state] (s_{10}) [right=of s_8] {$s_{10}$};
   \node[state] (s_7) [above right=of s_1] {$s_7$};
    \node[state] (s_9) [right=of s_7] {$s_9$};
   
    \path[->] 
    (s_0) edge  node {$\frac{sex=M}{P_{b}=4/5: \{\ell_3, \ell_4\}}$} (s_1) 
    (s_0) edge  node  [below left] {$\frac{sex=F}{P_{b}=1/5, {\{\ell_1, \ell_2\}}}$} (s_2)
    (s_1)  edge node [below] {$\frac{age=[40,50]}{P_{b}=1/4 : \{\ell_4\}}$} (s_8)
    (s_1)  edge node {$\frac{age=[30,40]}{P_{b}=3/4 : \{\ell_3\}}$} (s_7)
     (s_2)  edge node {$\frac{age=[30,40]}{P_{db} = 1 : \{\ell_1,\ell_2\}}$} (s_3)
    %  (s_3) edge node [above] {$\frac{response = 1}{P_{db} =1/2:  \{\ell_1\}}$} (s_5)
      (s_3) edge node{$\frac{P_{db} =1/2}{\{\ell_1\}}$} (s_5)
          (s_5) edge node{$\frac{response(\ell_1)=1}{P_{db} =1}$} (s'_5)
    % (s_3) edge node [below] {$\frac{P_{db} =1/2} {\{\ell_2\}}$} (s_6)
      (s_3) edge node [below] {$\frac{P_{db} =1/2} {\{\ell_2\}}$} (s_6)
          (s_6) edge node{$\frac{response(\ell_2) = 8}{P_{db}=1}$} (s'_6)
    %   (s_7) edge node [below] {$\frac{response = 3}{P_{db}=1:  \{\ell_3\}}$} (s_9)
       (s_7) edge node [above] {$\frac{response(\ell_3)=3}{P_{db}=1}$} (s_9)
    %  (s_8) edge node [below] {$\frac{response = 7}{P_{db}=1:  \{\ell_4\}}$} (s_{10})
       (s_8) edge node [above] {$\frac{response(\ell_4)= 7}{P_{db}=1}$} (s_{10})  
     ;
\end{tikzpicture}
\caption{\label{f2}DLTTS-B  for $B$'s capture of  responses} 
  \end{figure}

\vspace*{-1em}
\paragraph{Attacker $B$:} 
With DLTTS-B  (Figure~\ref{f2}), Attacker $B$ gets access to the responses
of employees, preference to  males of age 30 to 40.  ($P_b$ = probabilities
from $B$'s assigned objective.)

\lft 
Maximal probability thresholds  computed by $B$: Males $6/10$, Females $1/10$.
Note: The threshold $6/10$ for males is higher than the $3/16$  reported by $C$.

{The actual probabilities for all possible responses are: $Pr(response = 3|M) = 6/10$,  \\
  $Pr(response = 7|M) = 1/5, Pr(response = 1|F) = 1/10$ and $Pr(response = 8|F) = 1/10$. }

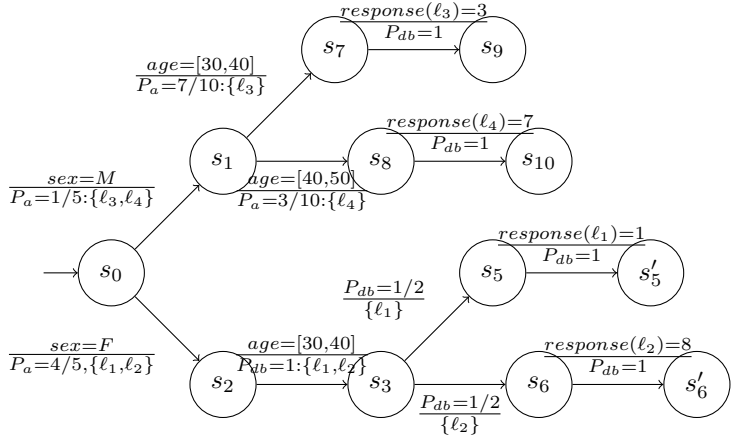
\begin{figure}%[h]
\begin{tikzpicture}[node distance=2.1cm,on grid,auto] 
   \node[state,initial, initial text ={}] (s_0)   {$s_0$}; 
   \node[state] (s_1) [above right=of s_0] {$s_1$};
   \node[state] (s_2) [below right=of s_0] {$s_2$}; 
    \node[state] (s_3) [right=of s_2] {$s_3$}; 
    \node[state] (s_5) [above right=of s_3] {$s_5$};
    \node[state] (s'_5) [right=of s_5] {$s'_5$};
    \node[state] (s_6) [right=of s_3] {$s_6$};
    \node[state] (s'_6) [right=of s_6] {$s'_6$};
   \node[state] (s_8) [right=of s_1] {$s_8$};
   \node[state] (s_{10}) [right=of s_8] {$s_{10}$};
   \node[state] (s_7) [above right=of s_1] {$s_7$};
    \node[state] (s_9) [right=of s_7] {$s_9$};
   
    \path[->] 
    (s_0) edge  node {$\frac{sex=M}{P_{a}=1/5: \{\ell_3, \ell_4\}}$} (s_1) 
    (s_0) edge  node  [below left] {$\frac{sex=F}{P_{a}=4/5, {\{\ell_1, \ell_2\}}}$} (s_2)
    (s_1)  edge node [below] {$\frac{age=[40,50]}{P_{a}= 3/10: \{\ell_4\}}$} (s_8)
    (s_1)  edge node {$\frac{age=[30,40]}{P_{a}=7/10 : \{\ell_3\}}$} (s_7)
    (s_2)  edge node {$\frac{age=[30,40]}{P_{db} = 1 : \{\ell_1,\ell_2\}}$} (s_3)
    (s_3) edge node {$\frac{P_{db}=1/2}{\{\ell_1\}}$} (s_5)
    (s_5) edge node {$\frac{response(\ell_1) =1}{P_{db}=1}$} (s'_5)
      (s_3) edge node [below] {$\frac{P_{db} ={1/2}} {\{\ell_2\}}$}
      (s_6)
    (s_6) edge node{$\frac{response(\ell_2) = 8}{P_{db}=1}$} (s'_6)
    (s_7) edge node [above] {$\frac{response(\ell_3) =3}{P_{db}=1}$} (s_9)
    (s_8) edge node [above] {$\frac{response(\ell_4)= 7}{P_{db}=1}$} (s_{10})  
    ;
\end{tikzpicture}
\caption{\label{f3}DLTTS-A  for $A$'s capture of responses} 
  \end{figure}

\vspace*{-1em}
\paragraph{Attacker $A$:} 
With DLTTS-A  (Figure~\ref{f3}) Attacker $A$ gets access to the responses of employees,
preference to  females of age 30 to 40.  ($P_a$ = probabilities from $A$'s assigned
objective.)

\lft 
Maximal probability thresholds computed by $A$: Males $7/50$, Females $2/5$. \\
Note: The threshold {$2/5$} for females is higher than the $3/16$
reported by $C$.

The actual probabilities for possible responses are:  $Pr(response = 3|M) = 7/50$, \\
 $Pr(response = 7|M) = 3/50, Pr(response = 1|F) = 2/5$ and $Pr(response = 8|F) = 2/5$.

\vspace*{2mm}
In view of our earlier remark,  we may deduce from these details, that: \par
 - $B$'s attack  succeeds at node $s_7$ ($response(\ell_3)$) and 
   node $s_8$ ($response(\ell_4)$), on DLTTS-B. \par 
 - $A$'s attack succeeds at node $s_5$ ($response(\ell_1)$)  and 
   node $s_6$ ($response(\ell_2)$), on DLTTS-A. 

 \vspace*{1.5mm}
 Such `defeats' for the Basic Analyser can be avoided if the administrator of  $D$ (and
 the oracle mechanism $\o$ in the DLTTS) implement a strategy for  better  protecting
 the access to the special values in the base $D$ (`responses', in this example). 
 We present such a strategy in  the following subsection. 

\vspace*{-1.5em}
\subsection{A Strategy for better 'Secrecy' Protection}~\label{Strategy}
\vspace*{-1.5em}

For any attacker $N$, and  a  DLTTS-N constructed by $N$  modeling his/her
query-runs on the given base $D$,  if $s$ is a node where the incoming transition
has a singleton label set  $\{\ell\}$, we shall denote by $Pr(s, l; N)$ the probability
computed at $s$ with $\succ$-priority transitions along the runs to $s$ from 
the initial state. We shall denote by $Max_{pr}(\ell; N)$ the maximum
of all these probabilities, taken over all such nodes $s$ on DLTTS-N. 

\vspace*{1mm} \lft
(1)  Let $s$ be a node on DLTTS-B, with a singleton $\{\ell\}$ labeling the  incoming
transition.  \\ 
\hspace*{2.2cm} IF \, $Max_{pr}(\ell; C)  <  Pr(s, \ell; B) $,  \\ 
\hspace*{1.5cm}  THEN {\em Switch OFF} the outgoing transition
                                                    {\em response(l)} at $s$.
        
\lft        
(2)  Do the same on  DLTTS-A, with respect to DLTTS-C.

 \vspace*{2mm}\lft
 {\bf Details:}

 \lft
 (i) \hspace*{2mm} We have: \, $3/16 = Max_{pr}(\ell_3; C) <  Pr(s_7, \ell_3; B) = 3/5$, \par 
    \hspace*{1cm} and  \;\; $3/16 = Max_{pr}(\ell_4; C) <  Pr(s_8, \ell_4; B) = 1/5$. 

 \lft 
 (i) \hspace*{2mm} We have:  \, $3/16 = Max_{pr}(\ell_1; C) < Pr(s_5, \ell_1; A) = 2/5$,  \par 
 \hspace*{1cm} and \;\;  $3/16 = Max_{pr}(\ell_2; C)  < Pr(s_6, \ell_2;  A) = 2/5$.   
  
   \vspace*{1mm}
   It follows  that the outgoing transition `response()' can be   switched OFF, if we apply
   the    the strategy above  at nodes $s_7, s_8$ on DLTTS-B, and at nodes $s_5, s_6$
   on DLTTS-A.   \hfill$\Box$
      
 \vspace*{2mm}
    We are in a position  now  to formulate an empirical strategy  for better  protecting access
    to the `special' values, for any database.  The formulation below is for {\em any}
    general   database  $D$  with a `column of protected values' (anonymized or not) that
    we   shall still refer   to as `responses', and {\em any} attacker $N$.
   It is assumed in this formulation that  the   administrator of the base  $D$ has made an
   appropriate choice for the `Basic Analyser' $C$. 
      
    \vspace*{1mm}\lft
    {\bf The Strategy:} 
    
    $\bullet$ Let $s$ be a node on a DLTTS-N  under construction by an Attacker $N$ for access
    to the  responses,  where  the incoming transition at $s$ on DLTTS-N has a singleton label
    ${\ell}$. 
     
     $\bullet$ Suppose  the probability $Pr(s, \ell; N)$ at $s$, computed  along the runs to
      $s$ with the supposed initial knowledge of $N$, and $\succ$-priority transitions all along, 
     satisfies the condition: \par
     \hspace*{2.4cm} $ Pr(s, \ell; N)) >  Max_{pr}(\ell; C)$.  \par%\vspace*{(-1mm}
     Then {\em switch OFF} the  outgoing transition $response(\ell)$ at the node $s$ on
     DLTTS-N.

%%%%%%%%%%%%%%%%%%%%%%%%%%%%%%%%%%%%
 \vspace*{-1.2em}
  \section{ {\large Related Work and Comments} }~\label{Conclude}
  \vspace*{-1.2em}

  Our work started with the observation that  databases   could be distributed over
  several `worlds', so querying  such bases   leads  in general to  answers which
  would also be distributed; and to the  distributed  answers  one could conceivably
  assign probability distributions  of relevance  to   the query.  It seemed  thus natural
  to view  the  probabilistic automata of Segala (\cite{Segala95a,Segala95b}),
  with  outputs, as an appropriate logical  structure for analyzing formally, 
  the evolution of distributed information  under  the transitions of these automata.
  Distributed  Transition   Systems (DTS) appeared a little   later, but most of  them
  had as objective  the behavioral  analysis of the  distributed   transitions, based on
  traces or  on   simulation/bisimulation.   Quasi- or pseudo-  or hemi- metrics,
  suitably defined,  turned out to be  essential for  the reasonings  employed,
  for instance,   as in~\cite{Fast2018,PTS2019,LBr-SystMetrics09}.   On the other hand,
  approaches  based on metrics  have been studied,  in particular in  \cite{dpMetrics2013},
  for  refining notions of differential privacy and of adjacency, for databases  in
  `standard formats' (numerical  or strings., all of the same dimension).  
  
  Our lookout for a metric based  vision   for privacy analysis has   been
  influenced  by many of these works,  although not  with the same objective.
  As the developments in this work show, our {\em syntax}-based   metric can
  almost directly handle data of  `mixed types':  they can be numbers or
  literals , but can  also be `anonymized'   as intervals or sets; they can also be
  taxonomically  related to each other on a  tree structure.

  As has been shown in Section~\ref{NewDefn}, the value-wise (partial) metric, 
  constructed in  Section~\ref{DataWise} on type compatible sets of data, has
  led us to a finer notion of $\epsilon$-distinguishabilty on  mecha\-nisms answering
  queries.  The practical application for the DLTTS vision that we have  presented in
  Section $8$ of the current paper,  is an addition to our earlier work~\cite{siva-etal-2022}. 
  Although rather simple, it must be sufficiently illustrative of how the   DLTTS  vision
  can be used. On the other hand, the syntactic developments presented in Section
  $8$  have certain similarities, in our opinion, with the semantic considerations
  presented in  \cite{ChenChu2008}. 

  We have not considered any notions of noisy channels perturbing   numerical
  data in the databases; but  it  is not difficult to extend the DLTTS setup -- and 
  the  mechanism $\M$ answering the queries -- of our work, to handle noise
  additions.  It will then be assumed  that the  internal  (saturation) procedure
  $\C$, at  every  state in  the DLTTS,  incorporates the  three well-known
  noise  adding  mechanisms:  the Laplace, Gauss, and  exponential mechanisms,  
  with the assumption that noise additions to  numerical values  is done
  in a {\em bounded} fashion -- as in e.g., \cite{BdLaplace2020}, so as to be
  from  a  finite  prescribed   domain  around the values. It will then be
  assumed  that the tuples  with noisy data are also  in  the base signature  $\E$. 
  The  notion  of $\epsilon$-local-indistinguishabilty between  tuples with
  noisy  data   can also be defined in such an extended setup.

  As part of future work, we hope to generalize the value-wise (partial) metric
  constructed in  Section~\ref{DataWise} of the current paper, by assigning
  different `weights' to the columns of the given base. That could be one of
  the techniques to  `disfavor'  the columns in the base that tend to be
  `noisier', or of lesser interest.  That would also offer the  possibility of  taking
  into   account   possible dependencies between some of the columns in the base.
  We   hope to deduce still finer notions of adjacency on  databases,   and  of
  $\epsilon$-distinguishabilty on query answering mechanisms, with such 
  a refinement.   As concerns  the  strategy for better secrecy protection, proposed
  in  Section~\ref{Strategy}, the crucial assumption is on the appropriate choice 
  of the `Basic Analyser' by the system administrator; it seems rather specific
  to the  context/example considered, and notions  like   `completeness'   or
  `soundness'  of  such a strategy may not be easily formalizable.  

    \section{Conclusion}
    \label{Conclusion}

   We have presented in this article the DLTTS model, whose goal is to capture the knowledge
   that a querier (considered in this work as an \emph{honest-but-curious} attacker)
   accumulates when querying a database containing private information, protected by simple
   privacy policies aiming to protect the values of some specific tuples.
   We show how DLTTS can be used as a core model by a database administrator to detect
   privacy breaches and show how it can be used to implement a simple strategy that
   consists in not answering any further queries.
%  \pagebreak

 %%%%%%%%%%%%%%%%%%%%%%%%%%%%%%%%%%%%

%  \pagebreak
  
 \vspace*{-1mm}
  \section{ {\large Appendix}}~\label{WP-Metric}
  \vspace*{-1em}
  
  Taxonomies are frequent in machine learning.  Data mining  and clustering  techniques
  employ reasonings based on measures of symmetry, or on metrics, depending  on the
  objective. The Wu-Palmer symmetry  measure on tree-structured   taxonomies is
  one among those in use; it is defined as follows (\cite{WuPalm1994}):
  Let $\T$ be a  given  taxonomy tree.   For any node $x$ on $\T$, define its depth
  $c_x$ as the number of nodes  from the root  to $x$ (both included), along the path
  from the root to $x$. For any pair $x, y$ of nodes  on $\T$, let $c_{x y}$ be the depth
  of the common ancestor of $x, y$ that is  {\em farthest } from the root.
  The Wu-Palmer symmetry  measure between the nodes  $x, y$  on $\T$ is then
  defined as WP$(x, y) = \frac{2 \, c_{x y}}{c_x + c_y}$.
  This measure, although considered satisfactory for many purposes, is known to have
  some disadvantages such as not being conform to semantics in several situations.

  What we are interested in, for the purposes of our current paper, is a   {\em metric}
  between the nodes of a taxonomy tree, which in addition will suit our semantic
  considerations.  This is the objective of our Lemma below. (A result that seems to be
  unknown, to our knowledge.) 

 % \vspace*{-1mm}
  \begin{lemma} % ~\label{wpmetric}
    On any taxonomy tree $\T$, the  binary function between its
    nodes defined by \,  $d_{wp}(x, y) = 1 - \frac{2 \, c_{x y}}{c_x + c_y}$ {\em (notation
    as above)} is a metric. 
  \end{lemma}

  \vspace*{-1mm}
   \lft {\em Proof}:
We drop the suffix $wp$ for this proof, and just write $d$.
  Clearly $d(x, y) = d(y, x)$; and $d(x, y) = 0$ if and only if $x = y$.  We only have to
  prove the  Triangle Inequality; i.e. show that   $d(x,z) \le d(x, y) + d(y,z)$ holds for
  any three nodes $x,y,z$ on $\T$.  A `configuration'  can be  typically represented
  in its `most  general form'  by the  diagram below.
  The boldface characters   $X, Y, Z, a, h$ in the diagram all stand   for the  {\em
    number of arcs}  on the corresponding paths.  So that, for the depths  of $x, y, z$,
  and  of their farthest common  ancestors on the tree, we get:
  
  \disp{$c_x = X + h + 1 ,  \; \; c_y = Y + h + a + 1,  \;\;  c_z = Z + h + a + 1$, \\ 
     $c_{xy} = h + 1, \;\;   c_{yz} = h + a + 1,  \;\;  c_{xz} = h + 1$}

  %\vspace*{-0.8em} \lft
  \lft
 The `$+1$' in these equalities is because the $X, Y, Z, a, h$ are the {\em number of 
 arcs} on the paths,  while the depths are the number of nodes. The $X, Y, Z, a, h$ must
 all be  integers $\ge 0$. For the Triangle Inequality on  the three nodes $x, y, z$ on $\T$,
 it suffices to prove the following two relations:
 \disp{$d(x, z) \le d(x, y) + d(y, z)$ \, and \,  $d(y, z) \le d(y, x) + d(x, z)$.}

% \vspace*{-0.6em} 
 \lft by showing that the following two algebraic inequalities hold:

\vspace*{1mm} 
\disp{\!(1) $1 - \frac{ 2*(h+1) }{ (X+Y+2*h+a+2) } + 1 - \frac{ 2*(h+a+1) }{ (Y+Z+2*h+2*a+2) }$
  $\ge$     $1 - \frac{  2*(h+1) }{ (X+Z+2*h+a+2) }   $ \\ \vspace*{1mm}
   \, (2) $1 - \frac{ 2*(h+1) }{ (X+Y+2*h+a+2) } + 1 - \frac{ 2*(h+1) }{ (X+Z+2*h+2*a+2) }$
   $\ge$     $1 - \frac{  2*(h+a+1) }{ (Y+Z+2*h+2*a+2) }   $}

 \lft The third  relation $d(x, y) \le  d(x, z) + d(z, y)$ is proved by just
exchanging the roles of $Y$ and $Z$ in the proof  of inequality (1).

\lft Inequality (1): We eliminate  the denominators (all  strictly positive), and  write
 it out as an inequality between two polynomials  $eq1, eq2$ on $X,Y,Z$, $h,a$, which
 must be satisfied for all their non-negative integer values:
 
\vspace*{1mm}
\lft $ eq1: (X+Y+2*h+a+2)*(Y+Z+2*h+2*a+2)*(X+Z+2*h+a+2)$ \par
\lft $ eq2:  (h+1)*(Y+Z+2*h+2*a+2)*(X+Z+2*h+a+2) $ \\ \hspace*{1.5cm}
                    $ +(h+a+1)*(X+Y+2*h+a+2)*(X+Z+2*h+a+2)$ \\ \hspace*{1.5cm}
                    $ - (h+1)*(X+Y+2*h+a+2)*(Y+Z+2*h+2*a+2)$  \\ 
$eq: eq1 - 2*eq2$.  \; We need to check:\, $eq \ge 0$ ? 

\vspace*{1mm}\lft
The equation $eq$ once expanded  (e.g., under {\em Maxima})  appears as:

\vspace*{1mm}
\disp{
$eq: YZ^2+XZ^2+aZ^2+Y^2Z+2XYZ+4hYZ+2aYZ+4YZ+X^2Z+4hXZ+2aXZ+4XZ+a^2Z+XY^2
               +4hY^2+aY^2+4Y^2+X^2Y+4hXY+2aXY+4XY+8h^2Y+8ahY+16hY+a^2Y+8aY+8Y$}
\lft
The coefficients are all positive, and inequality (1)  is proved. 

\vspace*{-1mm}
\begin{center}
 \includegraphics[scale=0.4]{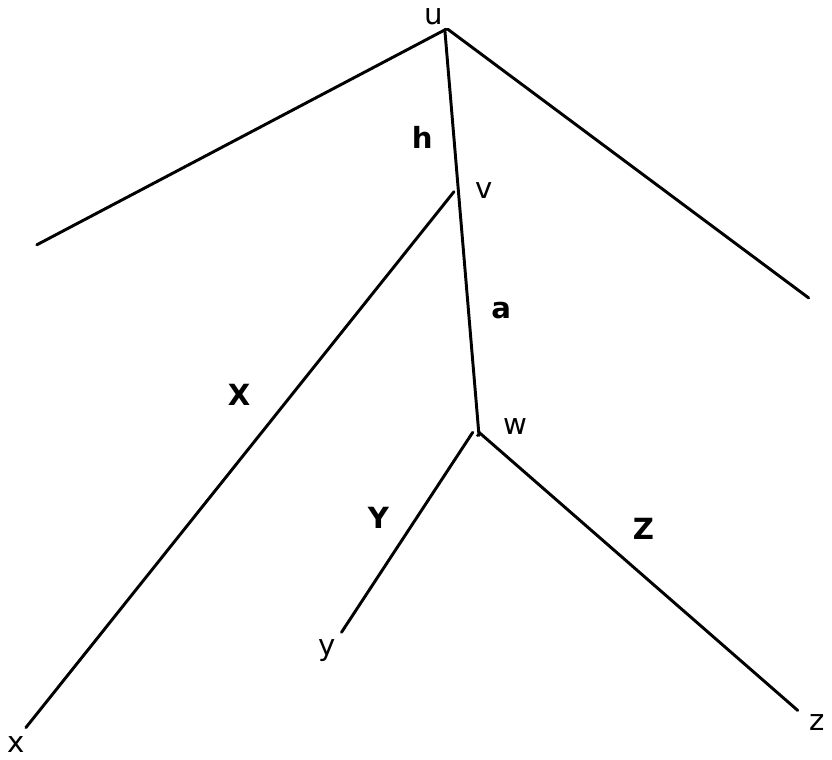}
  \end{center}
\vspace*{-4.9cm}

\vspace*{2mm}
\lft Inequality (2): We first define the following polynomial expressions: 

\vspace*{2mm}
\lft $eq3:  (X+Y+2*h+a+2) * (X+Z+2*h+a+2) * (Y+Z+2*h+2*a+2)$;

\lft $eq4:  (h+1) * (Y+Z+2*h+2*a+2) * ( 2*X+Y+Z+4*h+2*a+4)$;

\lft $eq5: (h+a+1) * (X+Y+2*h+a+2) * (X+Z+2*h+a+2)$;

\vspace*{2mm}
\lft If we set \, $eqn: eq3 + 2*eq5 - 2*eq4$, \,  we get 

\vspace*{2mm}
\lft $eqn:  - 2(h+1) * (Z+Y+2h+2a+2) * (Z+Y+2X+4h+2a+4) + \\ \hspace*{6mm}
           (Y+X+2h+a+2) * (Z+X+2h+a+2)(Z+Y+2h+2a+2) + \\ \hspace*{6mm}
             2 (h+a+1) * (Y+X+2h+a+2)* (Z+X+2h+a+2)$

 \lft Inequality (2) is proved by showing that $eqn$ remains non-negative
 for all  non-negative values of $X,Y,Z,h,a$; we expand $eqn$ (with {\em Maxima}), 
to  get:                  

 \vspace*{2mm}
  \lft $eqn$:
  $Y Z^2 + X Z^2 + a Z^2 + Y^2 Z + 2 XYZ + 4 hYZ + 6 aYZ + 4YZ + X^2 Z + 4 hXZ + 6 aXZ + 4 XZ
      + 8 ahZ + 5 a^2Z + 8 aZ + XY^2 + aY^2 + X^2 Y + 4 hXY + 6 aXY + 4XY + 8 ahY + 5 a^2Y 
      + 8 aY+ 4 hX^2 + 4 aX^2 + 4X^2 + 8 h^2X + 16 ahX + 16 hX + 8 a^2X
         + 16 aX + 8 X + 8 ah^2 + 12 a^2h + 16 ah + 4 a^3 + 12 a^2 + 8 a$

\vspace*{2mm}
\lft The coefficients are all positive, so we are done.  \hfill $\Box$. 


\begin{thebibliography}{99}

  \bibitem{siva-etal-2022}
  S.~Anantharaman, S.~Frittella, B.~Nguyen.
  \newblock ``Privacy Analysis with a Distributed Transition System and a Data-Wise Metric''
  \newblock In: Privacy in Statistical Databases,  PARIS, France, Lecture Notes in  Computer
  Science, Vol. PSD 2022 (LNCS 13643). Pp. 15-30, Springer, 09. 2022.
  
 \bibitem{BarthePOPL12}
  G.~Barthe, B.~K\"{o}pf, F.~ Olmedo, S.Z.~B\'eguelin.
  \newblock ``Probabilistic relational reasoning for differential privacy''.
  \newblock In: Proceedings of POPL, ACM (2012)
 
  \bibitem{BartheLICS20}
  G.~Barthe, R.~Chadha, V.~Jagannath,  A.~Prasad Sistla, M.~Viswanathan.
  \newblock   ``Deciding Differential Privacy for Programs with Finite Inputs
                  and  Outputs''.
 \newblock In: LICS'20: 35th Annual {ACM/IEEE} Symposium on Logic in Computer
 Science, Saarbr{\"{u}}cken, Germany, July 8-11, 2020.

\bibitem{Fast2018}     
  V.~Castiglioni, K.~Chatzikokolakis, C.~Palamidessi.
 \newblock ``A Logical Characterization of Differential Privacy via Behavioral Metrics''.
 \newblock In: Formal Aspects of Component Software (FACS), Pohang,
 South Korea. pp. 75--96, Oct. 2018. 
 
\bibitem{PTS2019}     
  V.~Castiglioni, M.~Loreti, S.~Tini.
  \newblock ``The metric linear-time branching-time spectrum on
                       nondeterministic probabilistic processes''.
  \newblock  In: Theoretical Comp. Science, Vol. 813:20--69, 2020. 

\bibitem{dpMetrics2013}
  K.~Chatzikokolakis, M.~Andr\'es, N.~Bordenabe, C.~Palamidessi.
  \newblock  ``Broadening the Scope of Differential Privacy Using Metrics''.
  \newblock In: Privacy Enhancing Technologies Symposium (PETS), 
  Bloomington, IND (US), pp. 82--102, 2013,

\bibitem{ChenChu2008}
  Y.~Chen, W.~W.~Chu.
  \newblock ``Database Security Protection via Cokllaborative Inference Detection''.
  \newblock In: IEEE Transactions on Knowledge and Data Engineering,
     20(8): 1013-1027 (2008). 
  
\bibitem{LBr-SystMetrics09}
  L.~de Alfaro, M.~Faella, M.~Stoelinga.
  \newblock ``Linear and Branching System Metrics''. 
  \newblock In: IEEE Trans. on  Software Engineering, Vol. 35(2):258--273, 2009.

  \bibitem{Quasi-Id}
    T. Dalenius.
   \newblock  ``Findig a Needle in Haystack'' (or `Identifying Anonymous Census Records')
   \newblock In: J. of Official Statistics, Vol. 2 No. 3, pp. 329--336, 1986. 
    
  \bibitem{Dwork2006}
 C.~Dwork.
 \newblock ``Differential privacy''.
 \newblock  In: Proceedings of ICALP 2006. LNCS (Springer--Verlag), 
                        Vol. 4052, pp. 1--12, 2006.

 \bibitem{Dwork2014}
  C.~Dwork. A.~Roth.
  \newblock ``The Algorithmic Foundations of Differential Privacy''.
  \newblock In: Found. Trends Theor. Comput. Sci., Vol. 9:3-4, pp. 211--407, 2014. 
  
 \bibitem{BdLaplace2020}
N.~Holohan, S.~Antonatos, S.~Braghin, P.~M.~Aonghusa.
\newblock ``The Bounded Laplace Mechanism in Differential Privacy''.
 \newblock In: Journal of Privacy and Confidentiality (Proc. TPDP 2018), Vol. 10 (1), 2020. 

 \bibitem{Segala95c}
 R.~Segala.
 \newblock ``Modeling and Verification of Randomized Distributed
           Real-Time   Systems''.
 \newblock  Ph.D. thesis, MIT (1995).
           
 \bibitem{Segala95a}          
 R.~Segala.
 \newblock ``A compositional trace-based semantics for probabilistic automata''.
 \newblock  In: Proc. CONCUR'95, 1995, pp. 234--248.

\bibitem{Segala95b}
  R.~Segala, N.A. Lynch.
  \newblock ``Probabilistic simulations for probabilistic processes''.
  \newblock In: Nord. J. Comput. 2(2):250--273, 1995.
  
\bibitem{Warner65}
  Stanley L. Warner.
  \newblock ``Randomized Response: A Survey Technique for Eliminating Evasive Answer Bias''
  \newblock  In: Journal of the American Statistical Association
    Vol. 60(309), pp. 63--69, 1965. 
  
 \bibitem{WuPalm1994}
   Z.~Wu, M.~Palmer.
 \newblock ``Verb Semantics and Lexical selection''.
 \newblock  In: Proc. 32nd Annual meeting of the Associations for Comp. Linguistics,
    pp 133-138. 1994.
  
\end{thebibliography}
 \end{document}